\documentclass[twocolumn,notitlepage,aps,pra,11pt,superscriptaddress]{revtex4-1}
\usepackage[utf8]{inputenc}
\usepackage{braket}
\usepackage{nicefrac}
\usepackage{enumitem}
\usepackage{amsthm,amsmath,amssymb,amsfonts}
\usepackage{mathtools}
\usepackage{latexsym}
\usepackage{graphicx}  
\usepackage{bm}        
\usepackage{verbatim}
\usepackage{soul}
\usepackage{mathpazo}
\usepackage{tikz}
\usepackage{hyperref}
\usepackage{multirow}
\usepackage{array}
\newcolumntype{P}[1]{>{\centering\arraybackslash}p{#1}}

\usepackage{pgfplots}
\usetikzlibrary{plotmarks}

\newlength\figureheight
\newlength\figurewidth
\pgfplotsset{compat=1.14, 
             every axis/.append style={
                    tick label style={/pgf/number format/fixed}
                    }}

\usepackage{lipsum}
\usepackage{titlesec}

\titlespacing\section{0pt}{12pt plus 4pt minus 2pt}{0pt plus 2pt minus 2pt}
\titlespacing\subsection{0pt}{12pt plus 4pt minus 2pt}{0pt plus 2pt minus 2pt}
\titlespacing\subsubsection{0pt}{12pt plus 4pt minus 2pt}{0pt plus 2pt minus 2pt}
\begin{document}
\title{Comparing neural network based decoders for the surface code}
\author{Savvas Varsamopoulos}\thanks{svarsamo@gmail.com}
\author{Koen Bertels}
\author{Carmen Garcia Almudever}
\affiliation{Quantum and Computer Engineering, Delft University of Technology, Mekelweg 4, 2628 CD Delft, The Netherlands}

\begin{abstract}
Matching algorithms can be used for identifying errors in quantum systems, being the most famous the Blossom algorithm. Recent works have shown that small distance quantum error correction codes can be efficiently decoded by employing machine learning techniques based on neural networks (NN). Various NN-based decoders have been proposed to enhance the decoding performance and the decoding time. Their implementation differs in how the decoding is performed, at logical or physical level, as well as in several neural network related parameters. In this work, we implement and compare two NN-based decoders, a low level decoder and a high level decoder, and study how different NN parameters affect their decoding performance and execution time. Crucial parameters such as the size of the training dataset, the structure and the type of the neural network, and the learning rate used during training are discussed. After performing this comparison, we conclude that the high level decoder based on a Recurrent NN shows a better balance between decoding performance and execution time and it is much easier to train. We then test its decoding performance for different code distances, probability datasets and under the depolarizing and circuit error models.
\end{abstract}
\maketitle

Quantum computers are a promising solution to a class of complex problems that classical supercomputers cannot currently solve or require an immensely large amount of time to solve. However, since quantum computing is still in its early stages, classical computers are still the driving force, using the prototypes of quantum computers as accelerators for specific applications.

In the recent past, there is an increasing dominance of heterogeneous, multi-core architectures with multiple processors. In such architectures, a classical core processor interacts with different co-processors such as Graphics Processing Units (GPUs), Field Programmable Gate Arrays (FPGAs), Tensor Processing Units (TPUs) and in this case a quantum processor. Such a quantum processor requires both classical and quantum computing components, because it needs a lot of monitoring and control from the classical part. Typically, when developing such an architecture, one has to develop a full stack going from algorithms up to the chip implementation.
    
\begin{figure}[htb]
\centering
\scalebox{0.7}{
\includegraphics[width=\columnwidth]{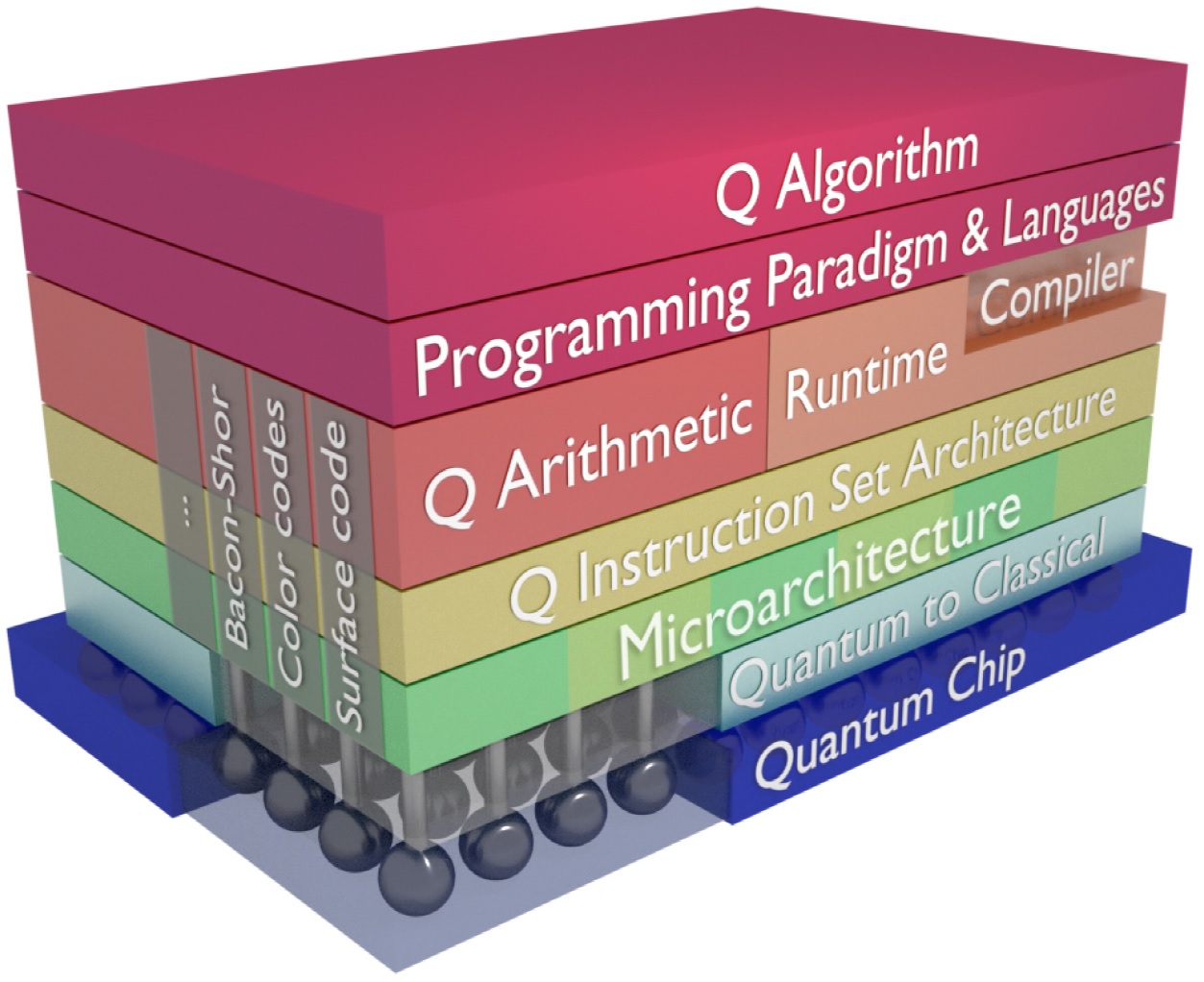}}
\caption{Overview of the quantum computer system stack}
\label{fig:stack}
\end{figure}   

Figure \ref{fig:stack} provides an overview of the quantum system stack consisting of the following layers \cite{fu2016heterogeneous}: 
\begin{itemize}[label=$\diamond$]
    \item The top layers involve the quantum algorithms alongside the language constructs and compilers that are required to generate a series of instructions that belong to the Quantum Instruction Set Architecture (QISA).
    
    \item The micro-architecture layer translates these instructions into pulses to operate in the quantum chip. These pulses are sent through the quantum to classical interface.

    \item As can be seen by the $3^{rd}$ dimension of Figure \ref{fig:stack}, quantum error correction (QEC) is a key part when building a fault-tolerant quantum accelerator and it affects several layers of the stack, including the micro-architecture.
\end{itemize}

\begin{figure*}[htb]
\centering
\scalebox{2.0}{
\includegraphics[width=\columnwidth]{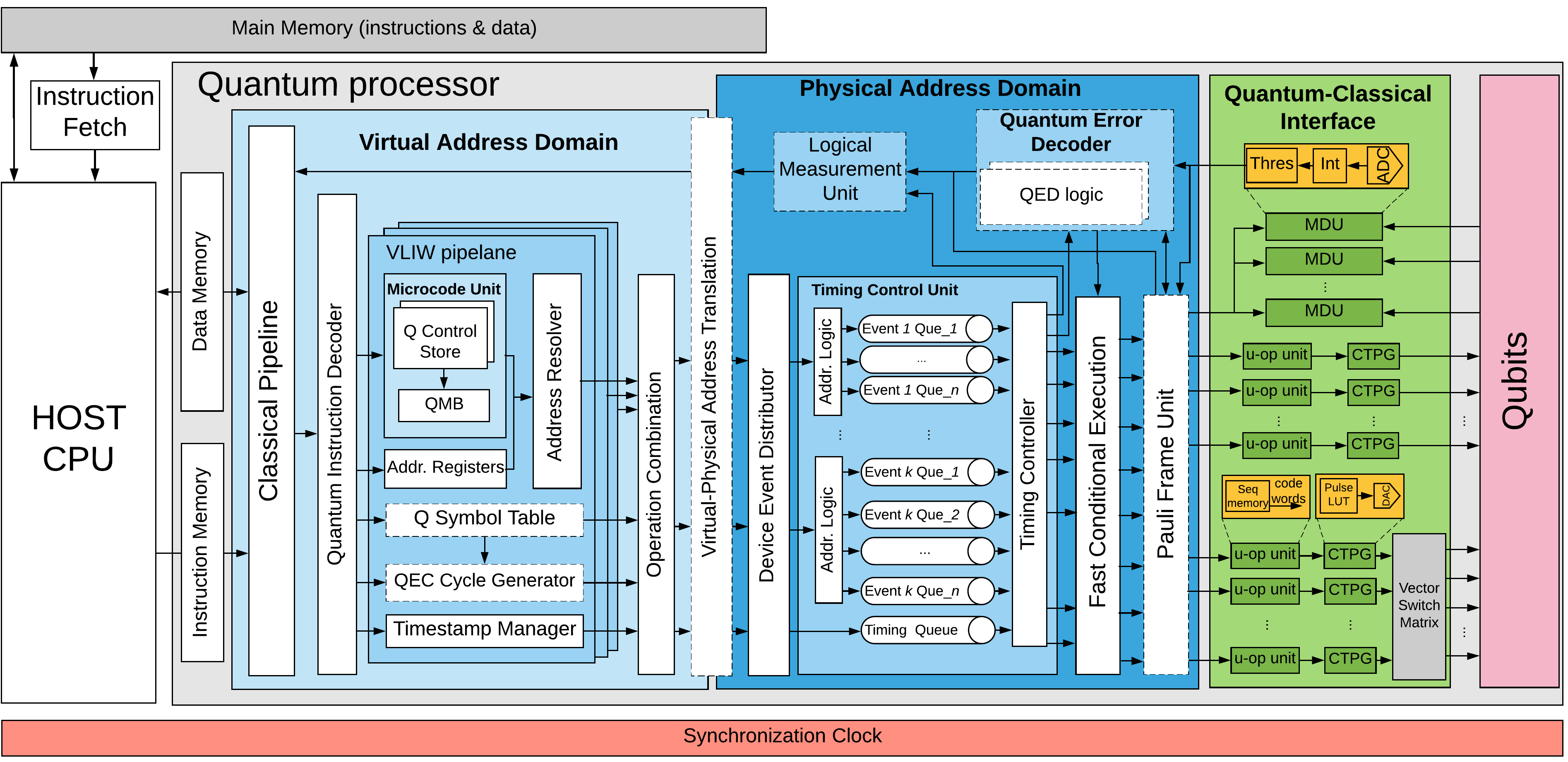}}
\caption{Overview of the quantum micro-architecture \cite{fu2017experimental}}
\label{fig:uarch_envision}
\end{figure*}

As shown in Figure \ref{fig:uarch_envision}, different micro-architectural blocks are required to keep track of the errors and identify the location and type of these errors such as the Decoding block and the Pauli frame unit. In this work, we focus on the Decoder logic block of Figure \ref{fig:uarch_envision} that is part of the QEC process, as we will explain in the next paragraphs.

Constant active quantum error correction is regarded as necessary in order to perform reliable quantum computation and storage due to the unreliable nature of current quantum technology. Qubits inadvertently interact with their environment even when no operation is applied, forcing their state to change (decohere). Moreover, application of imperfect quantum gates results in the introduction of errors in the quantum system. Quantum error correction is the mechanism that reverses these errors and restores the state to the desired one. 

Quantum error correction involves an encoding and a decoding process. Encoding is used to enhance protection against errors by employing more resources (qubits). In this paper, we limit ourselves to \textbf{\textit{surface code}} encoding \cite{gottesman, Kitaev2003}. Decoding is the process that is used to identify the location and type of error that occurred. As part of quantum error correction, decoding has a limited time budget that is determined by the time of a single round of error correction. In the case that the decoding time exceeds the time of quantum error correction, either the quantum operations are stalled or a backlog of inputs to the decoding algorithm is created \cite{toms_simulation_paper}. Many classical decoding algorithms have been proposed with the most widely used being the \textbf{\textit{Blossom decoder}} \cite{edmonds}. Blossom has been shown to reach high decoding accuracy, but its execution time scales polynomially with the number of qubits \cite{dklp}, which can be problematic for large quantum systems needed to solve complex problems. However, there are optimized versions of Blossom for topological codes, that report linear scaling with the number of qubits \cite{fowler_opt_comp} and even a parallel version stating that the average processing time per detection round is constant independent of the size of the system \cite{fowler_O1}. Also, there exist other decoders like the union find decoder that is presented in \cite{Nickerson}, which report an almost linear scaling of the execution time as the quantum system increases linearly.
In addition, Blossom's execution time also depends on the physical error rate. Blossom performs a Minimum Weight Perfect Matching (MWPM) on a graph that is created based on the amount of errors that have been generated and detected. Then, higher error rates lead to bigger graphs and longer execution time. In this case, the execution time scales polynomially with the physical error rate.

An alternative to classical decoders is to use a neural network for identifying errors. They exhibit constant execution time with the physical error rate and their execution time scales linearly with the linear increase of the number of qubits. Also, they have been proven to provide better decoding performance than many classical decoding algorithms \cite{Savvas,Baireuther,Torlai,Krastanov,Breuckmann,Chamberland2018,Maskara_2018}. Note that, most of neural network based decoders use the neural network as a probability distribution, however there exist decoders that create an exact mapping between belief propagation and deep neural networks, as presented in \cite{Liu}.

Neural network based decoder implementations differ in how the neural network performs the decoding, the type and structure of the network, the amount of samples used for training and many other aspects. Therefore, their decoding performance and their reported execution time also differs depending on the implementation choices that were made. 

The contributions of this paper can be summarized as follows:

\begin{enumerate}
\item So far, there is no thorough comparison made between different neural network based decoders under the same conditions. In this paper, we have implemented two different NN-based decoders, namely the high level and low level decoder, and analyzed their decoding performance and execution time while exploring different NN parameters. We show that the high level decoder exhibits constant execution time regardless of the physical error rate and scales linearly with the linear increase of the qubits in the quantum system, while able to reach at least equivalent decoding performance to the Blossom decoder (baseline).
\item Usually when NN-based decoders are designed, achieving the highest decoding performance is the main goal, however, in this work we emphasize the importance of small execution time, due to the limited available time budget of quantum error correction. Therefore, the neural network based decoders implemented in this paper should present a good balance between decoding performance and execution time.
\item We analyze how the choice of the dataset affects the training of neural network based decoders. We show that sampling and training at the different physical error rates that the decoder is being evaluated, leads to higher decoding performance than sampling and training to a single physical error rate and evaluating the decoder at a vastly different error rate.

\end{enumerate}

The rest of the paper is organized as follows: in section~\ref{sec:QEC} we provide a brief introduction to quantum error correction. In section~\ref{sec:Des_NNs}, we explain how the different NN-based decoders, as found in literature, are implemented. In section~\ref{sec:Impl_param}, many parameters of the neural network based decoders are discussed. In section~\ref{sec:Results}, we provide the results with the best neural network based decoder for the different error models. Finally, in section~\ref{sec:Conclusions}, we draw our conclusions about this research.

\section{Quantum error correction}
\label{sec:QEC}
Similar to classical error correction, quantum error correction encodes a set of unreliable \textbf{\textit{physical}} qubits to a more reliable qubit, known as \textbf{\textit{logical}} qubit. 

Various quantum error correcting codes have been developed so far, but in this work we only consider the surface code \cite{gottesman,Kitaev2003,freedman2001projective,bravyi1998quantum,bombin2007optimal}, one of the most promising QEC codes. The \textbf{\textit{surface code}} is a topological stabilizer code that has a simple structure, local interactions between qubits and is proven to have high tolerance against errors \cite{dklp,Raussendorf, fowler_high_thres,Wang_high,fowler_clas_proc,Bombin_2009,Bombin_2011,bravyi1998quantum,fowler_sc}. It is usually defined as a planar lattice of qubits over two dimensions.

\begin{figure}[htb]
\centering 
\scalebox{0.8}{
\includegraphics[width=\columnwidth]{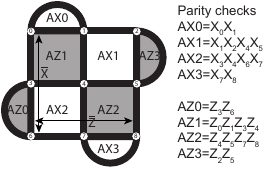}}
\caption{Rotated surface code with code distance 3. Data qubits are enumerated from 0 to 8. X-type ancilla are in the center of the white tiles and Z-type ancilla are in the center of grey tiles}\label{fig:parity_checks}
\end{figure}

In the surface code, a logical qubit consists of physical qubits that store quantum information, known as \textbf{\textit{data}} qubits, and physical qubits that are used to detect errors in the logical qubit through their measurement, known as ancillary or \textbf{\textit{ancilla}} qubits (see Figure \ref{fig:parity_checks}). A logical qubit is defined by its \textbf{\textit{logical operators}} ($\bar{X}$,$\bar{Z}$) that define how the logical state of the qubit can be changed. Any operator of the form $X^{\otimes n}$ or $Z^{\otimes n}$ that creates a chain that span both boundaries of the same type can be regarded as a logical operator, with $n$ being the number of data qubits that are included in the logical operator.

An important feature of the surface code is the code distance. \textbf{\textit{Code distance}}, (\textbf{\textit{d}}), describes the degree of protection against errors. More accurately, is the minimum number of physical operations required to change the logical state \cite{gottesman, lidar}. In surface code, the degree of errors (d.o.e.) that can be successfully corrected, is calculated according to the following equation:

\begin{equation}
	\text{d.o.e.} = \bigg \lfloor \frac{d-1}{2} \bigg \rfloor
\end{equation}

The smallest surface code created is known as the rotated surface code \cite{Horsman2012} and it is presented in Figure \ref{fig:parity_checks}. It consists of 9 data qubits placed at the corners of the square tiles and 8 ancilla qubits placed inside the square and semi-circle tiles. Each ancilla qubit can interact with its neighboring 4 (square tile) or 2 (semi-circle tile) data qubits. 

\begin{figure}[htb]
\centering 
\scalebox{0.8}{
\includegraphics[width=\columnwidth]{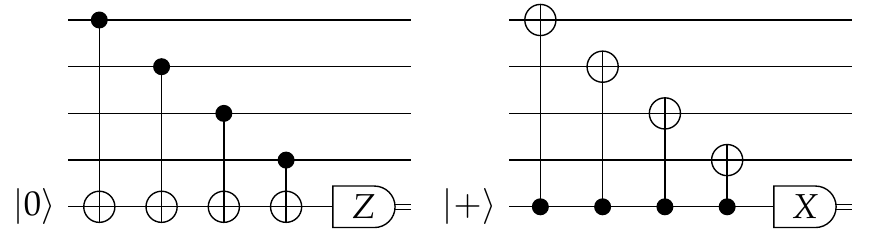}}
\caption{Syndrome extraction circuit for individual Z-type (left) and X-type (right) ancilla, with the ancilla placed in the bottom}\label{fig:SCcycle}
\end{figure}

As mentioned, ancilla qubits are used to detect errors in the data qubits. Although quantum errors are continuous, the measurement outcome of each ancilla discretizes quantum errors into bit-flip (X) and phase-flip (Z) errors, that can be detected by Z-type ancilla and X-type ancilla, respectively. The circuit that is used to collect the ancilla measurements for the surface code is known as \textbf{\textit{syndrome extraction circuit}}. It is presented in Figure \ref{fig:SCcycle} and it signifies one round of error correction. It includes the preparation of the ancilla in the appropriate state, followed by 4 (2) CNOT gates that entangle the ancilla qubit with its 4 (2) neighboring data qubits and then the measurement of the ancilla qubit in the appropriate basis. The measurement result of the ancilla is a binary value, which is calculated as the parity between the state of the data qubits connected to it. Each ancilla performs a parity-check of the form of $X^{\otimes 4}$/$Z^{\otimes 4}$ (square tile) and $X^{\otimes 2}$/$Z^{\otimes 2}$ (semi-circle tile), as presented in Figure \ref{fig:SCcycle}. When the state of the data qubits involved in a parity-check has not changed, then the parity-check will return the same value as in the previous error correction cycle. In the case where the state of an odd number of data qubits involved in a parity-check is changed compared to the previous error correction cycle, the parity-check will return a different value than the one of the previous cycle ($0\leftrightarrow1$). The change in a parity-check in consecutive error correction cycles is known as a \textbf{\textit{detection event}}.

Note that the parity-checks are used to identify errors in the data qubits without having to measure the data qubits explicitly and collapse their state. The state of the ancilla qubit at the end of every parity-check is collapsed through the ancilla measurement, but is initialized once more in the beginning of the next error correction cycle \cite{Versluis}.

The parity-checks must conform to the following rules: i) must commute with each other, ii) must anti-commute with errors and iii) must commute with the logical operators. An example of these parity-checks for a d=3 rotated surface code is presented in Figure \ref{fig:parity_checks}. The notation $X_i$ or $Z_i$ refers to the $i^{th}$ data qubit used in a given parity-check.

Gathering all measurement outcomes, forms the \textbf{\textit{error syndrome}}. Surface code can be decoded by collecting the ancilla measurements out of one or multiple rounds of error correction and providing them to a decoding algorithm that identifies the errors and outputs data qubit corrections.

\begin{figure}[htb]
\centering 
\scalebox{0.87}{
\includegraphics[width=\columnwidth]{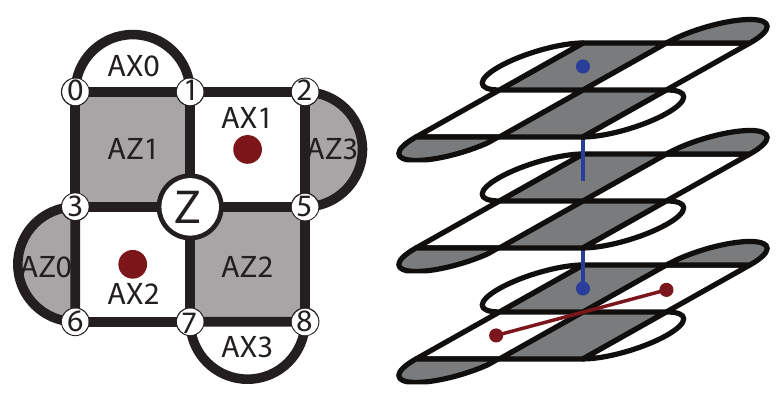}}
\caption{Rotated surface code with code distance 3. Left: Phase-flip (Z) error at data qubit 4, which causes two detection events (shown in red). Right: Three consecutive rounds of error correction. The red dots indicate detection events that arise from a data qubit error and the blue dots indicate detection events that arise from a measurement error}\label{fig:2D_and_3D}
\end{figure}

A single error on a data qubit will be signified by a pair of neighboring parity-checks changing value from the previous error correction cycle. In the case where an error occurs at the sides of the lattice, only one parity-check will provide information about the error. Multiple data qubit errors that occur near each other, form one dimensional chains of errors which create only two detection events located at the endpoints of the chains (see Figure \ref{fig:2D_and_3D} on the left side and the red line on the right side). On the other hand, a measurement error, which is an error during the measurement itself, is described as a chain between the same parity-check over multiple error correction cycles (see the blue line in Figure \ref{fig:2D_and_3D} on the right side). This blue line represents an alternating pattern of the measurement values (0-1-0 or 1-0-1) coming from the same parity-check for consecutive error correction cycles. If such a pattern is identified and is not correlated with a data qubit error, then it is considered a measurement error, so no corrections should be applied. Therefore, to properly distinguish between data and measurement errors, multiple error correction cycles need to be run before corrections are proposed.

There exist decoding algorithms that can decode efficiently the surface code, however optimal decoding is a NP-hard problem \cite{NPhard}. For example, maximum likelihood decoding (MLD) searches for the most probable error that produced the error syndrome, whereas Blossom searches for the least amount of errors that produced the error syndrome \cite{fowler2009high, fowler_opt_comp}. MLD has a significantly higher decoding performance than Blossom. However,it has an exponentially increasing execution time while the code distance increases linearly. For that reason, approximate versions of MLD have been developed like the \cite{Bravyi_Suchara_Vargo}, which reports a running time of $O(n\chi^3)$, where $\chi$ is a parameter that controls the approximation precision.

Although the Blossom decoder reaches lower decoding performance than MLD, the execution time of the optimized version exhibits linear scaling with the number of qubits. Furthermore, note that there exists a parallel version of Blossom described in \cite{fowler_O1}, that claims constant execution time regardless of the size of the system. Therefore, there is a clear trade-off between decoding performance and execution time, which is a key aspect of the decoder as there is a limited time budget for error correction.

The time budget for decoding is calculated based on the time required by the quantum operations involved in an error correction cycle. Therefore, the time budget varies based on the type of quantum technology, the choice of quantum error correcting code and the way quantum operations are performed. For example, in the case of superconducting qubits for a d=3 rotated surface code reported in \cite{toms_simulation_paper, Versluis}, the time budget for an error correction cycle is calculated to be \textasciitilde 700nsec, making most decoders unusable for near-term experiments based on these parameters. Furthermore, if noisy error syndrome measurements are assumed, then $d$ error correction cycles are required to provide the necessary information to the decoder, so in this scenario \textasciitilde 2.1$\mu$sec will be the upper limit for the time budget of decoding. However, as quantum technology matures and the fidelity of quantum operations increase, the time budget for QEC will also change.

An alternative decoding approach is to use neural networks to assist or perform the decoding procedure, since neural networks provide fast and constant execution time, while maintaining high application performance. In this paper, we are going to discuss decoders that include neural networks and compare them to each other and to the un-optimized version of the Blossom algorithm as described in \cite{kolmogorov}. In the following two sections we describe the implementation details of the NN-based decoders and the tuning of certain NN parameters.

\section{Neural network based decoders}
\label{sec:Des_NNs}
Artificial neural networks (ANN) have been shown to reach high application performance and constant execution time after being trained on a set of data generated by the application. ANN is a collection of weighted interconnected nodes that can transmit signals to each other. The receiving node processes the incoming signal and sends the processed result to its connected node(s). In this work, we focus on two types of neural networks known as \textbf{\textit{Feed-forward neural networks (FFNN)}} and \textbf{\textit{Recurrent neural networks (RNN)}}. In the case of RNNs, we used \textbf{\textit{Long Short-Term Memory (LSTM) cells}}. Since it was easy to generate input data and their corresponding output from the simulations, we used \textbf{\textit{supervised learning}} to train the neural network. Finally, we used the \textbf{\textit{mean squared error rate}} as a cost function and the \textbf{\textit{Rectified Linear Unit (ReLU)}} as the activation function.

Neural network based decoders for quantum error correcting codes have been recently proposed \cite{Savvas, Torlai, Krastanov, Baireuther, Chamberland2018, Maskara_2018}. There are two categories in which they can be divided: i) decoders that search for exact corrections at the physical level and ii) decoders that search for corrections that restore the logical state. We are going to refer to the former ones as \textbf{\textit{low level decoders}} (lld) \cite{Torlai, Krastanov} and the latter ones as \textbf{\textit{high level decoders}} (hld) \cite{Savvas, Baireuther, Chamberland2018,  Maskara_2018}. 

In this paper, we implement both decoders and compare them to each other and with the un-optimized Blossom decoder. The comparison is in terms of the decoding performance and the execution time. In the next sections, each decoder implementation is explained and the differences between them are presented.

\subsection{Inputs/Outputs}
Low level decoders take as input the error syndrome and produce as output an error probability distribution for each data qubit based on the observed syndrome. Therefore, a prediction is made that attempts to correct exactly all physical errors that have occurred.

High level decoders take as input the error syndrome and produce as output an error probability for the logical state of the logical qubit. Based on this scheme, the neural network does not have to predict corrections for all data qubits, rather just for the state of the logical qubit, which makes the prediction simpler. This is due to the fact that there are only 4 options as potential logical errors, $\bar{I}, \bar{X}, \bar{Z}, \bar{Y}$, compared to the case of the low level decoder where the output is equivalent to the number of data qubits. Moreover, trying to correctly predict each physical error requires a level of high granularity which is not necessary for error correcting codes like the surface code.

\subsection{Sampling and training process}
During the sampling process, multiple error correction cycles are run and the corresponding inputs and outputs for each decoder are stored. Due to the degenerate nature of the surface code, the same error syndrome might be produced by different sets of errors. Therefore, we need to keep track of the frequency of occurrence of each set of errors that provide the same error syndrome. 

For the low level decoder, based on these frequencies, we create an error probability distribution for each data qubit based on the observed error syndrome. For the high level decoder, based on these frequencies, we create an error probability distribution for each logical state based on the observed error syndrome.

When sampling is terminated, we train the neural network to map all stored inputs to their corresponding outputs. Training is terminated when the neural network is able to correctly predict at least 99\% of the training inputs. Further information about the training process are provided in section ~\ref{sec:Impl_param}.

\subsection{Implementation details}
\begin{figure}[htb]
\centering
\includegraphics[width=\columnwidth]{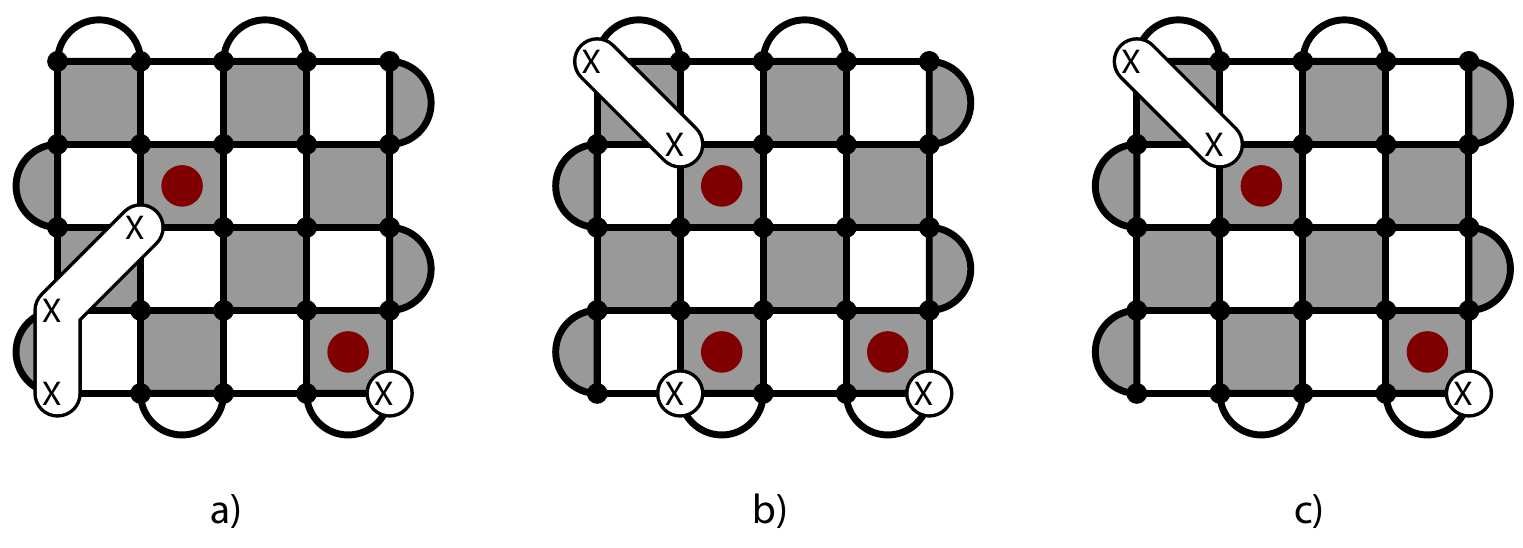}
\caption{Description of the decoding process of the low level decoder for a d=5 rotated surface code. (a) Observed error syndrome shown in red dots and bit-flip errors on physical data qubits shown with X on top of them. (b) \textbf{Invalid} data qubits corrections and the corresponding error syndrome. (c) \textbf{Valid} data qubits corrections and the corresponding error syndrome}\label{fig:nnd}
\end{figure}

Implementations of low level decoders typically include a single neural network. To obtain the predicted corrections for the low level decoder, we sample from the probability distribution that corresponds to the observed error syndrome for each data qubit, and predict whether a correction should be applied at each data qubit. However, this prediction needs to be verified before being used as a correction, because the proposed corrections must generate the same error syndrome as the one that was observed. Otherwise, the corrections are not valid (see Figure \ref{fig:nnd}a and b), since the decoder is predicting corrections for a different error syndrome than the one that was observed. In such a case the decoding performance will decrease significantly. Only when the two error syndromes match, the predictions are used as corrections on the data qubits (see Figure \ref{fig:nnd}a and c). If the observed syndrome does not match the syndrome obtained from the predicted corrections, then the predictions must be re-evaluated by re-sampling from the probability distribution. This re-evaluation step makes the decoding time non-constant, which can be a big disadvantage. There are ways to minimize the average amount of re-evaluations, however this is highly influenced by the physical error rate, the code distance and the strategy of re-sampling.

In Figure \ref{fig:nnd}, the decoding procedure of the low level decoder is described with an example. On \ref{fig:nnd}a, we present an observed error syndrome shown in red dots and the bit-flip errors on physical data qubits (shown with X on top of them) that created that syndrome. On \ref{fig:nnd}b, the decoder predicts a set of corrections on physical data qubits and the error syndrome resulting from these corrections is compared against the observed error syndrome. As can be seen from \ref{fig:nnd}a and \ref{fig:nnd}b, the two error syndromes do not match, therefore the predicted corrections are deemed \textbf{invalid}. On \ref{fig:nnd}c, the decoder predicts a different set of corrections and the corresponding error syndrome to these corrections is compared against the observed error syndrome. In the case of \ref{fig:nnd}a and \ref{fig:nnd}c, the predicted error syndrome matches the observed one, therefore the corrections are deemed \textbf{valid}.

\begin{figure}[htb]
\centering 
\includegraphics[width=\columnwidth]{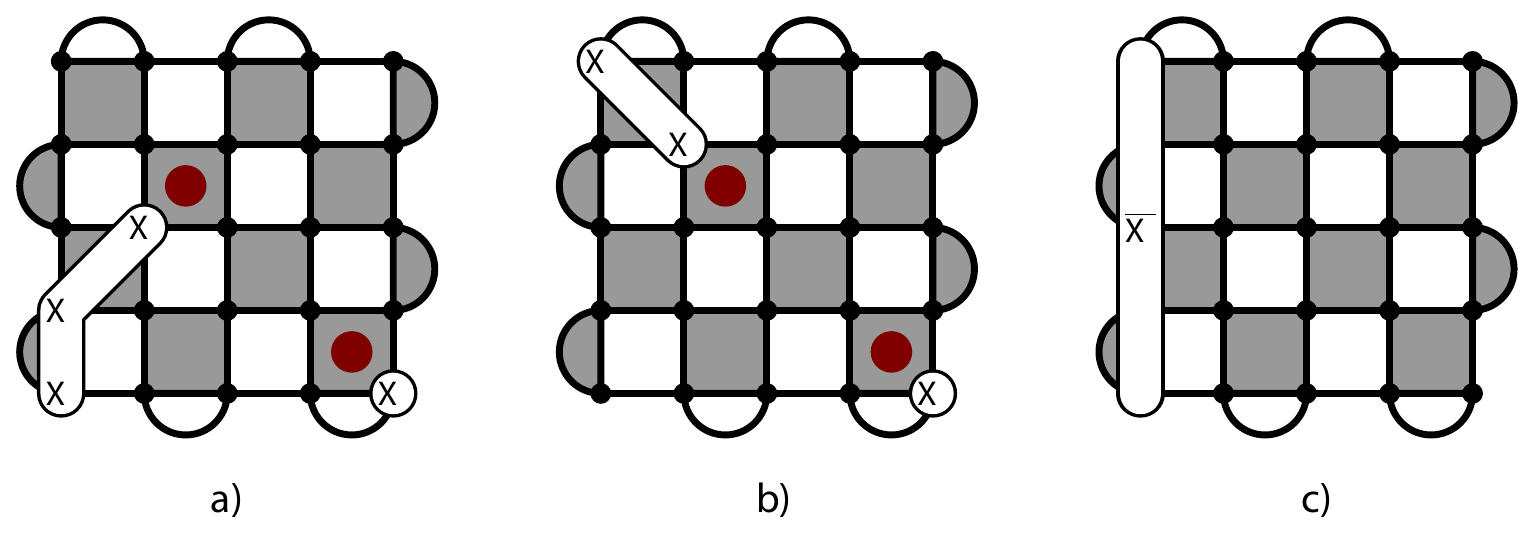}
\caption{Description of the decoding process of the high level decoder for a d=5 rotated surface code. (a) Observed error syndrome shown in red dots and bit-flip errors on physical data qubits shown with X on top of them. (b) Corrections proposed by the simple decoder for the observed error syndrome. (c) Additional corrections in the form of the $\bar{X}$ logical operator to cancel the logical error generated from the proposed corrections of the simple decoder}\label{fig:pfd}
\end{figure}

Implementations of high level decoders typically involve two decoding modules that work together to achieve high speed and high level of decoding performance. Either both decoding modules can be neural networks \cite{Baireuther} or one can be a neural network and the other one a non-neural network module\cite{Savvas, Chamberland2018}. The non-neural network module of the latter design will only receive the error syndrome out of the last error correction cycle and predict a set of corrections. In our previous experimentation \cite{Savvas}, this module was called \textbf{\textit{simple decoder}}. The corrections proposed by the simple decoder do not need to exactly match the errors that occurred, as long as the corrections correspond to the observed error syndrome (valid corrections). The other module which in both cases is a neural network, should be trained to receive the error syndromes out of all error correction cycles and predict whether the corrections that are going to be proposed by the simple decoder are going to lead to a logical error or not. In that case, the neural network outputs extra corrections, which are the appropriate logical operator that erases the logical error. The output of both modules is combined and any logical error created by the corrections of the simple decoder will be canceled due to the added corrections of the neural network (see Figure \ref{fig:pfd}). 

Furthermore, the simple decoder is purposely designed in the simplest way in order to remain fast, regardless of the quality of proposed corrections. By adding the simple decoder alongside the neural network, the corrections can be given at one step and the execution time of the decoder remains small, since both modules are fast and operate in parallel.

In Figure \ref{fig:pfd}, the decoding procedure of the high level decoder is described with an example. On \ref{fig:pfd}a, we present an observed error syndrome shown in red dots and the bit-flip errors on physical data qubits (shown with X on top of them) that created that syndrome. On \ref{fig:pfd}b, we present the decoding of the simple decoder. The simple decoder receives the last error syndrome of the decoding procedure and proposes corrections on physical qubits by creating chains between each detection event and the nearest boundary of the same type as the error. In Figure \ref{fig:pfd}b, the corrections on the physical qubits are shown with X on top of them, indicating the way that the simple decoder functions. The simple decoder corrections are always deemed \textbf{valid}, due to the fact that the predicted and observed error syndrome always match based on the construction of the simple decoder. In the case of Figure \ref{fig:pfd}a-b, the proposed corrections of the simple decoder are going to lead to an $\bar{X}$ logical error, therefore we use the neural network to identify this case and propose the application of the $\bar{X}$ logical operator as additional corrections to the simple decoder corrections, as presented in \ref{fig:pfd}c.

\section{Implementation parameters}
\label{sec:Impl_param}
In this paper, we compare both decoder designs. To achieve that, we implement the high level and low level decoder and test them under the same conditions.  We investigate how different implementation parameters affect the following metrics:

\begin{enumerate}
\item \textbf{The decoding performance:} it  indicates the accuracy of the algorithm during the decoding process. The typical way that decoding performance is evaluated is through lifetime simulations. In lifetime simulations, multiple error correction cycles are run and decoding is applied in frequent windows. Depending on the error model, a single error correction cycle might be enough to successfully decode, as in the case of perfect error syndrome measurements (window $=$ 1 cycle), or multiple error correction cycles might be required, as in the case of imperfect error syndrome measurements (window $=$ d cycles). When the lifetime simulations are stopped, the decoding performance is evaluated as the ratio of the number of logical errors found over the number of windows run until the simulations are stopped. 

\item \textbf{The execution time:} it is the time that the decoder needs to perform the decoding after being trained. It is calculated as the difference between the time when the decoder receives the first error syndrome of the decoding window and the time when it provides the output. For the low level decoder, the execution time will be the inference time of the neural network. For the high level decoder, the execution time will be the determined as the maximum time between the execution time of the simple decoder and the inference time of the neural network.
\end{enumerate}

These decoders were tested for two error models, the depolarizing error model and the circuit noise model. The \textbf{\textit{depolarizing error model}} assigns X,Z,Y errors with equal probability $\nicefrac{p}{3}$, known as depolarizing noise, only on the data qubits. No errors are inserted on the ancilla qubits and perfect parity-check measurements are used. Therefore, only a single cycle of error correction is required to find all errors. The \textbf{\textit{circuit noise model}} assigns depolarizing noise on the data qubits and the ancilla qubits. Furthermore, each single-qubit gate is assumed perfect but is followed by depolarizing noise with probability $\nicefrac{p}{3}$ and each two-qubit gate is assumed perfect but is followed by a two-bit depolarizing map where each two-bit Pauli has probability $\nicefrac{p}{15}$, except the error-free case, which has a probability of $1-p$. Depolarizing noise is also used at the preparation of a state and the measurement operation with probability $p$, resulting in the wrong prepared state or a measurement error, respectively. An important assumption is that the error probability of a data qubit error is equal to the probability of a measurement error, therefore $d$ cycles of error correction are deemed enough to decode properly.

\subsection{Choosing the training dataset}

The first step when designing a neural network based decoder is gathering data that will be used as the training dataset. The best dataset for a neural network based decoder is the dataset that achieves the highest decoding performance. Naively, one could suggest that including all possible error syndromes, would lead to the best decoding performance, however, as the code distance increases (size of the system increases), including all error syndrome becomes infeasible. As the code distance increases, the size of the space including all potential errors gets exponentially large as shown in Table \ref{table:datasets_depol} (state space). Therefore, we need to include as little but as diverse as possible error syndromes, which will provide the maximum amount of generalization. The size of the training dataset that we used for each code distance is also presented in Table \ref{table:datasets_depol} (training dataset). As shown, by employing such a technique, it seems impossible to continue beyond d=7 for the circuit noise model. For that distance, we gather error syndromes out of 10 error correction cycles and each error syndrome contains 48 ancilla qubits. Therefore, the full space that needs to be explored is 3.1x$10^{144}$. For the depolarizing error model, the highest code distance that we were able to decode efficiently was d=9. A potential candidate to assist in overcoming this scalability challenge and go to higher code distances is a distributed decoding strategy as found in \cite{Ni} and \cite{varsamopoulos2019decoding}.

The idea of using a distributed technique is to avoid processing all the error information from the whole code all-together. By limiting the amount of error information that is being processed at all times, the error syndrome space is always limited, thus making it easier to scale to larger code distances. In \cite{Ni}, a neural network decoder that is designed based on the Renormalization Group (RG) technique is explained. In RG decoding, the code is initially divided into small overlapping regions. Then, the error information out of these regions is combined to represent the error information of larger regions. This process of updating the error information while moving from a smaller region to a larger one is continued until the whole code ends up being one single region. In \cite{varsamopoulos2019decoding}, a similar approach is described. The main difference is that the code is always divided into specific size of overlapping regions and the information out of these small regions is forwarded to the whole code at once. Therefore, both techniques rely on dividing the code into overlapping regions and then updating the error information of each region in order to decode the whole code. Based on such a scheme, the maximum number of error information, which is translated into the maximum number of inputs of the neural network decoder, is always capped to a value that is much smaller compared to the case of the non-distributed approach.

\begin{table}[ht]
\centering
\caption{Dataset sizes for the depolarizing and the circuit noise error model}\label{table:datasets_depol}
\begin{tabular}{ | P{1.5cm} | P{3.1cm} | P{2.9cm} |} \hline
\multirow{2}{*}{}{Code} & State space & Training dataset  \\
distance & Depolarizing/ & Depolarizing/ \\ 
& Circuit & Circuit \\ \hline
\text{d=3} & 256\text{ / }4.3x$10^{9}$ & 256\text{ / }5x$10^{5}$  \\ \hline
\text{d=5} & 1.6x$10^{7}$\text{ / }4.3x$10^{9}$ & 2x$10^{5}$\text{ / }2x$10^{6}$   \\ \hline
\text{d=7} & 2.8x$10^{14}$\text{ / }3.1x$10^{144}$ & 3x$10^{6}$\text{ / }2x$10^{7}$  \\ \hline
\text{d=9} & 1.2x$10^{24}$\text{ / }9.7x$10^{288}$ & 2x$10^{7}$\text{ / }-   \\ \hline
\end{tabular}
\end{table}

In our previous experimentation\cite{Savvas}, we showed that sampling at a single physical error rate that always produces the fewest amount of corrections, is enough to decode small distance rotated surface codes with a decent level of decoding performance. This concept of always choosing the fewer amount of corrections is similar to the Minimum Weight Perfect Matching that Blossom algorithm uses. After sampling and training the neural network at a single physical probability, the decoder is tested against a large variety of physical error rates and its decoding performance is observed. We call this approach, the \textbf{\textit{single probability dataset}} approach, because we create only one dataset based on a single physical error rate and test it against many. Using the single probability dataset approach to decode various physical error probabilities is not optimal, because when sampling at low physical error rates, less diverse samples are collected, therefore the dataset is not diverse enough to correctly generalize to unknown training inputs.

The single probability approach is realistic for an experiment, since in an experiment there is only one error probability that the quantum system operates and at that probability the sampling, training and testing of the decoder will occur. However, this is not a good strategy for testing the decoding performance over a wide range of error probabilities. This is due to the degenerate nature of the surface code, since different sets of errors generate the same error syndrome. One set of errors is more probable when the physical error rate is small and another when it is high. When training the neural network, only one of these sets of errors, and always the same, is going to be selected for a given error syndrome regardless of the physical error rate being tested. Therefore, training a neural network based decoder in one physical error rate and testing its decoding performance in a widely different physical error rate might lead to poor decoding performance. The main benefit of this approach lies in the fact that only a single neural network has to be trained and used to evaluate the decoding performance for all the physical error rates that are tested. In the single probability dataset approach, the set with the fewer errors is always selected, because this set is more probable for the range of physical error rates that we are interested in.

To achieve a better decoding performance, we created multiple datasets that were obtained by sampling at various physical error rates and trained a different neural network at each physical error rate that was sampled. We call this approach, the \textbf{\textit{multiple probabilities datasets}} approach. Each dedicated training dataset that was created by sampling at a specific physical error probability is used to test the decoding performance at that same physical error probability and the probabilities close to that. By sampling, training and testing the performance for the same physical error rate, the decoder has the most relevant information to perform the task of decoding.

As we mentioned, depending on the sampling probability (physical error rate), different error syndromes will be more frequent than others. We chose to include the most frequent error syndromes in the training dataset. Since the objective of the decoding performance is to reach at least equivalent decoding performance to Blossom (baseline) with the smallest possible dataset, we increase the dataset size until it reaches Blossom's performance. Note that we do not claim to find the optimal training dataset to maximize the decoding performance, rather have a good balance between decoding performance and execution time. As we mentioned, we take the Blossom's performance as a baseline since it is the most popular decoder. Due to the multiple probabilities datasets, we train each dataset and evaluate the decoding performance for that probability and the ones close to it. 

The results of using a single probability dataset versus multiple probabilities datasets are shown in Figures \ref{fig:depol_sing_vs_mult} and \ref{fig:circ_sing_vs_mult} and will be discussed in Section 5.

\subsection{Structure of the neural network}
While investigating the size of a dataset, some preliminary investigation of the structure has been done, however only after the dataset is defined, the structure in terms of layers and nodes is explored in depth.

\begin{figure}[htb]
\centering 
\includegraphics[width=\columnwidth]{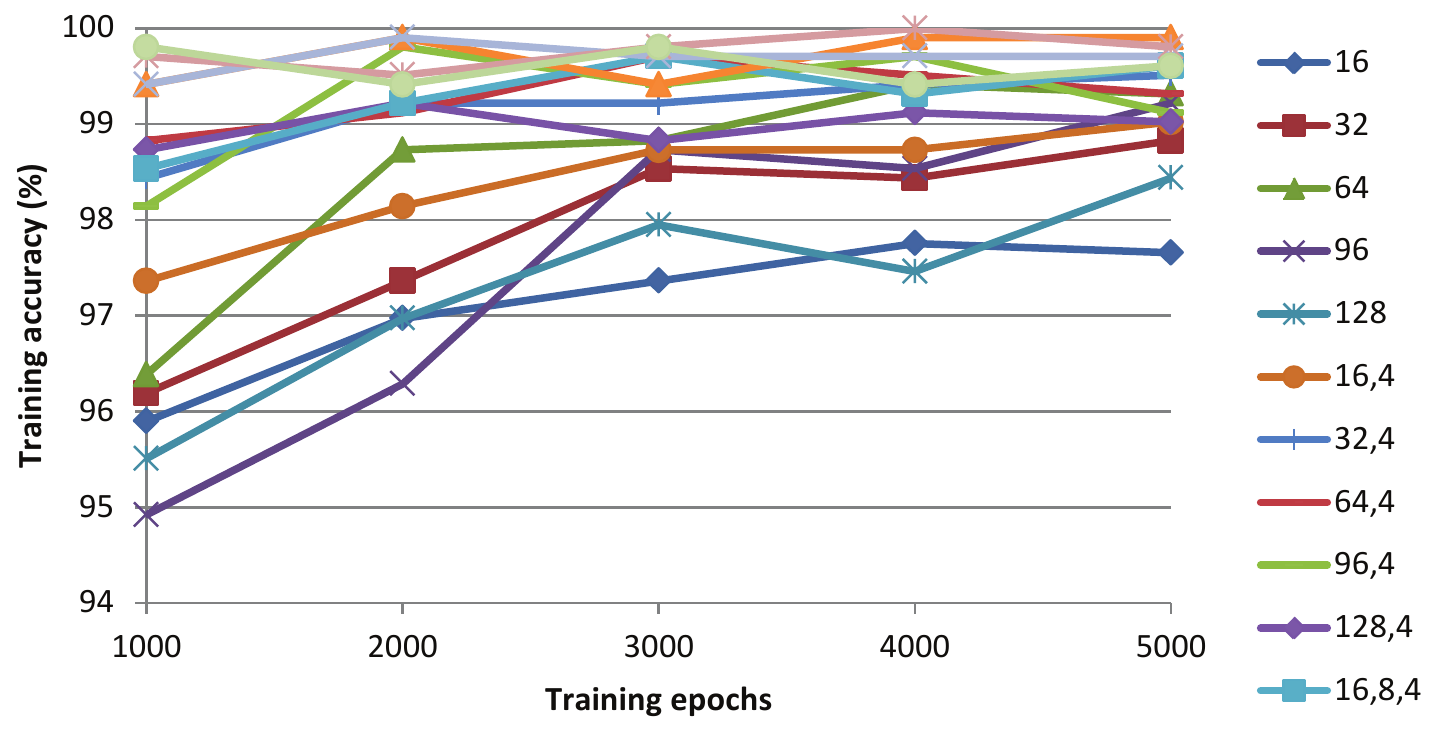}
\caption{Different configurations of layers and nodes for the high level decoder using a RNN (d=3, depolarizing error model). The nodes of the tested hidden layers are presented in the legend}\label{fig:training_acc}
\end{figure}

Initially, different configurations of layers and nodes have been tested for: i) distance 3 and 5 surface code, ii) the high- and low level decoder, iii) RNN and FFNN, and iv) the depolarizing error model. In our configurations, the number of nodes of the last hidden layer is selected to be equal to the number of output nodes. The rest of the hidden layers were selected to have decreasing number of nodes going from the first to the last layer. As an example, in Figure \ref{fig:training_acc} we present the investigation of multiple configurations of nodes and layers for the d=3 high level decoder using a RNN. Training stops at 5000 training epochs, since a good indication of the training accuracy achieved is evident by that point. Then, the one that reached the highest training accuracy was selected (in this case we chose the 16, 4). We continued training it till at least the same decoding performance as Blossom was reached.

In addition, we observed through this experimentation that the main factors that affect the configuration of the neural network are the size of the training dataset, the similarity between the training samples and the type of neural network. For instance, we found that the number of layers selected for training are affected more by the samples, e.g. the similarity of the input samples, and less by the size of the training dataset. 

\begin{figure}[htb]
\centering 
\includegraphics[width=\columnwidth]{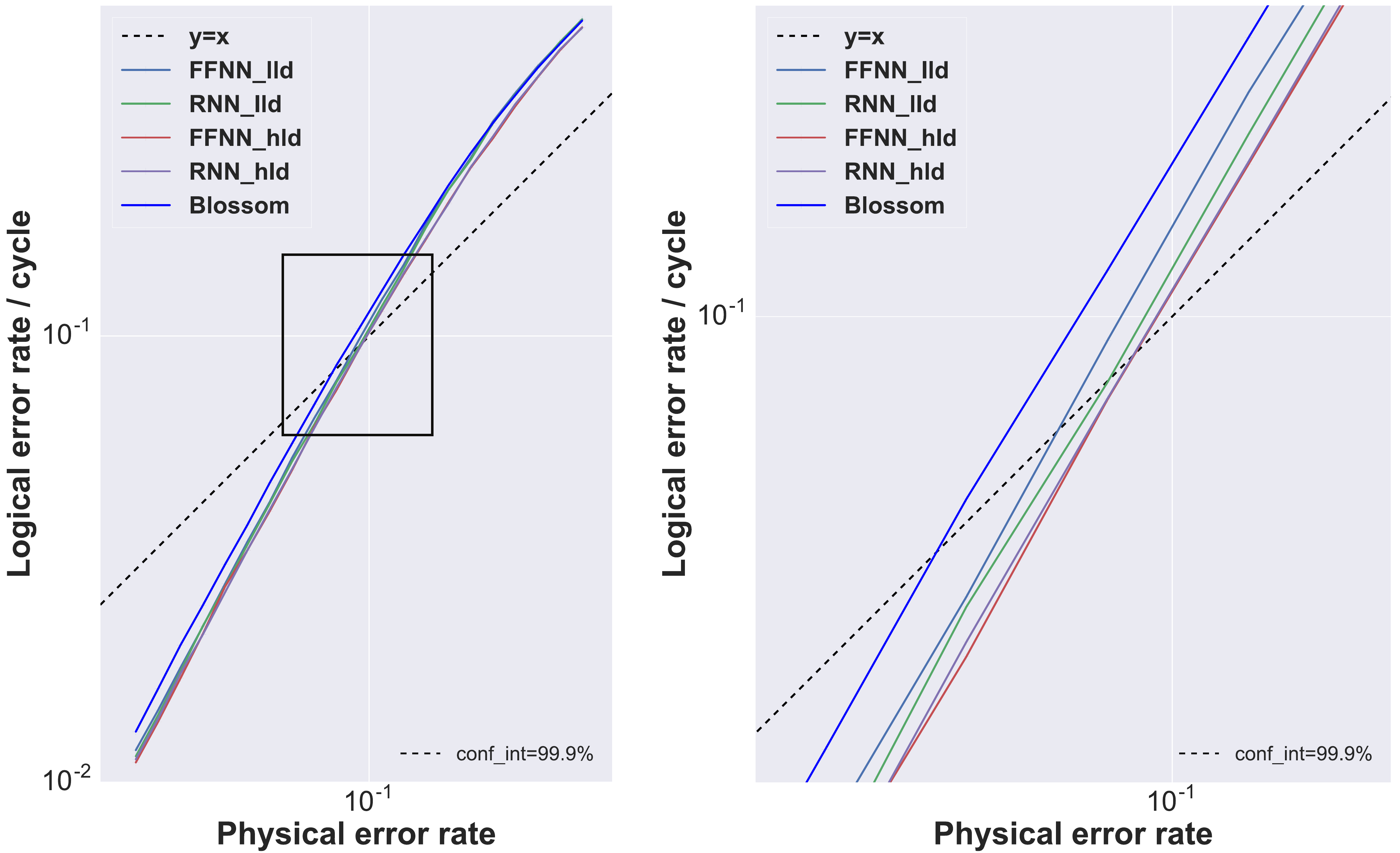}
\caption{Left: Comparison of decoding performance between Blossom algorithm, low level decoder and high level decoder for the d=3 rotated surface code for the depolarizing error model. Right: Zoomed in at the region defined by the square}\label{fig:d_3_nn_vs_sd}
\end{figure}

As we just mentioned, after selecting the configuration that showed the highest training accuracy we continue training it. In Figure \ref{fig:d_3_nn_vs_sd}, the decoding performance of the high level and low level decoder with FFNN and RNN and Blossom are shown. For even the small decoding problem of d=3, the more sophisticated Recurrent neural network seems to slightly outperform the Feed-forward neural network. This can also be seen in Table \ref{table:pseudo_d_3}, where the \textbf{\textit{pseudo-threshold}} values for d=3 and d=5 are shown. The pseudo-threshold is defined as the highest physical error rate that the quantum device should operate in order for error correction to be beneficial for a given code distance. Then, the highest the pseudo-threshold, the highest the decoding performance. The pseudo-threshold values for all decoders investigated in Figure \ref{fig:d_3_nn_vs_sd} can be found as the points of intersection between the decoder curve and the black dashed line, which represents the points where the physical error probability is equal to the logical error probability ($y=x$).

\begin{table}[ht]
\caption{Pseudo-threshold values for the tested decoders ($d$=3 and $d$=5) under depolarizing error model}\label{table:pseudo_d_3}
\centering
\begin{tabular}{ | P{2cm} | P{2.7cm} | P{2.7cm} |} \hline
\multirow{2}{*}{\text{Decoder}} & \text{Pseudo-threshold} & \text{Pseudo-threshold}  \\
& \text{d=3} & \text{d=5} \\ \hline
\text{FFNN lld} & 0.0911 & 0.1124\\ \hline
\text{RNN lld} & 0.0949 & 0.1172\\ \hline
\text{FFNN hld} & 0.0970 & 0.1219\\ \hline
\text{RNN hld} & 0.0970 & 0.1233\\ \hline
Blossom & 0.0825 & 0.1037\\ \hline
\end{tabular}
\end{table}

Note that, the difference between the decoding performance of the RNN and FFNN is even more obvious at larger code distances and for the circuit noise model, where the RNN naturally fits better due to its temporal dynamic behavior. Moreover, training of the FFNN becomes much harder compared to the RNN as the size of the dataset increases, making the experimentation with FFNN even more difficult.

Another observation from Figure \ref{fig:d_3_nn_vs_sd} and Table \ref{table:pseudo_d_3} is that the high level decoder is outperforming the low level decoder. Although there are ways to increase the decoding performance of the latter, mainly by re-designing the repetition step to find the valid corrections in less repetitions, we found no merit in doing so, since the decoding performance would still be similar to the high level decoder's and the repetition step would still not be eliminated.

\subsection{Training process}
\subsubsection{Batch size}
Training in batches instead of the whole dataset at once, can be beneficial for the training accuracy and training time. By training in batches, the weights of the neural network are updated multiple times per training iteration, which typically leads to faster convergence. We used batches of 1000 or 10000 samples, based on the size of the training dataset.

\subsubsection{Learning rate}
Another important parameter of training that can directly affect the training accuracy and training time is the learning rate. The learning rate is the parameter that defines how big the updating steps will be for each weight at every training iteration. Larger learning rates in the beginning of training can lead the training process to a minimum faster during gradient descent, whereas smaller learning rates near the end of training can increase the training accuracy. Therefore, we devise a strategy of a step-wise decrease of the learning rate throughout the training process. If the training accuracy has not increased after a specified number of training iterations (e.g. 50), then the learning rate is decreased. The learning rates used range from 0.01 to 0.0001. 

\subsubsection{Generalization}
The training process should not only be focused on the correct prediction of known inputs, but also the correct prediction of inputs unknown to training, known as generalization. Without generalization, the neural network acts as a Look-Up Table (LUT), which will lead to sub-optimal behavior as the code distance increases. In order to achieve high level of generalization, we continue training until the closeness between the desired and predicted value up to the $3^{rd}$ decimal digit is higher than 95\% over all training samples.

\subsubsection{Training time and execution time}
Timing is a crucial aspect of decoding and in the case of neural network based decoders we need to minimize both the training time and the execution time as much as possible.

\textbf{Training time:} it is the time required by the neural network to adjust its weights in a way that the training inputs provide the corresponding outputs as provided by the training dataset and adequate generalization can be achieved. It is proportional to the size of the training dataset and the number of qubits. The number of qubits is increasing in a quadratic fashion, $2d^{2}-1$, and the selected size of the training dataset in our experimentation is increasing in an exponential way, $2^{d^{2}-1}$. Therefore, training time should increase exponentially while scaling up. 

However, the platform that the training occurs, affects the training time immensely, since training in one/multiple CPU(s) or one/multiple GPU(s) or a dedicated chip in hardware will result in widely different training times. The neural networks that were used to obtain the results in this work, required between half a day to 3 days, depending on the number of weights and the inputs/outputs of the neural network, on a CPU with 28 hyper thread cores at 2GHz with 384GB of memory. 

\begin{figure}[htb]
\centering 
\includegraphics[width=\columnwidth]{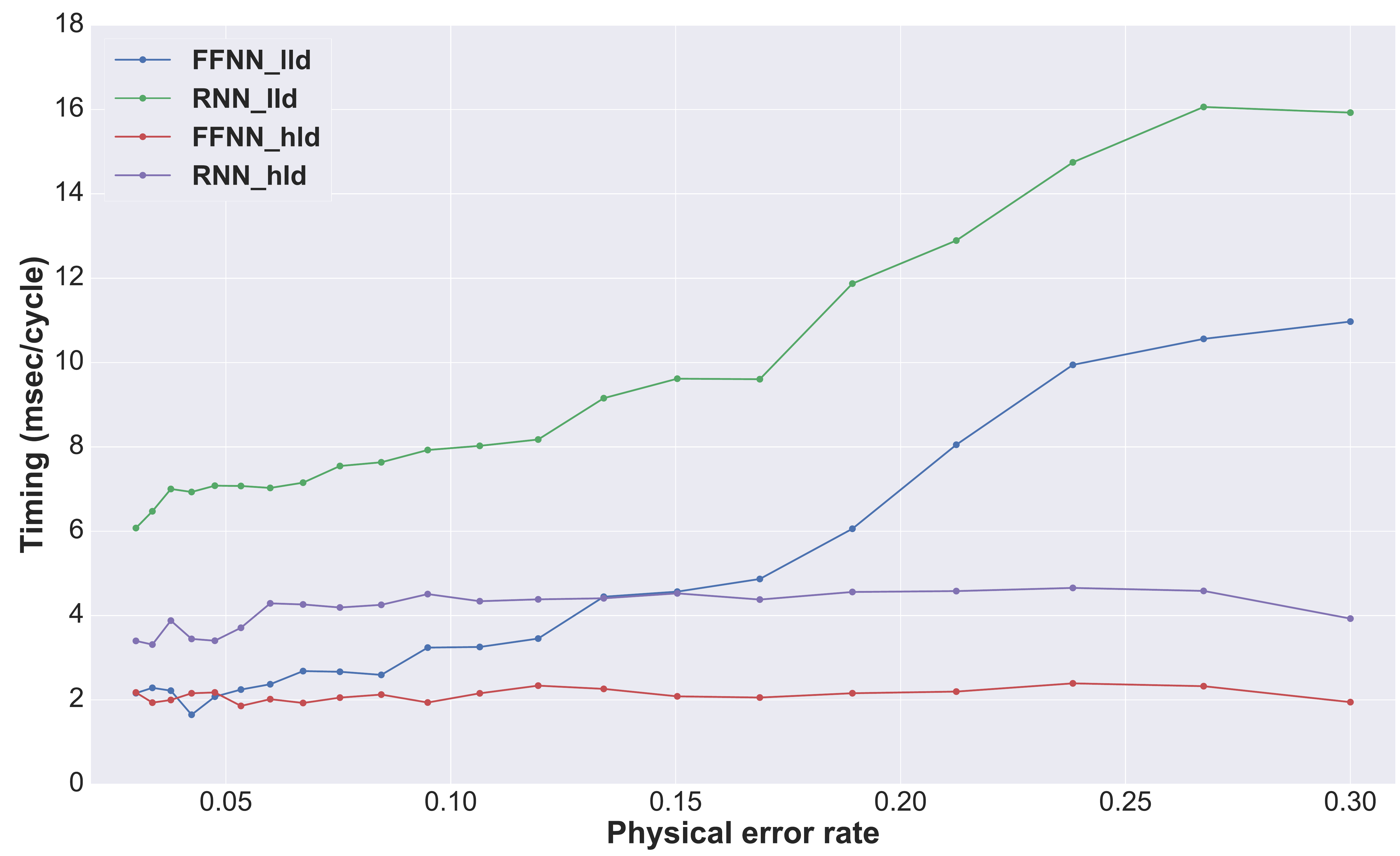}
\caption{Execution time for the high level decoder (hld) and the low level decoder (lld) for Feed-forward (FFNN) and Recurrent neural network (RNN) for d=3 rotated surface code for the depolarizing error model} \label{fig:timing_d_3}
\end{figure}

\textbf{Execution time:} In Figure \ref{fig:timing_d_3}, we present the constant and non-constant execution time with the physical error rate for the high level decoder and the low level decoder, respectively. The low level decoder has to repeat its predictions before it provides a valid set of corrections which makes the execution time non-constant. With careful design of the repetition step, the average number of predictions can decrease, however the execution time will remain non-constant. 

Note that results shown in Figure \ref{fig:timing_d_3} are only to illustrate the behavior of the execution time of the two decoders, therefore the time scale is not representative. In order to obtain a realistic time estimate, a hardware implementation should be made which has not been performed by this or any other group yet. 

As seen in Figure \ref{fig:timing_d_3}, the RNN has a higher execution time compared to the FFNN. This is due to the RNN typically using more weights compared to the FFNN. 

Although the execution time of the high level decoders is constant with the error rate, it appears to increase linearly with the code distance. In Table \ref{table:timing_depol}, we provide the calculated average time of decoding a surface code cycle under depolarizing noise for all distances tested with the high level decoder with RNNs.

\begin{table}[ht]
\caption{Average time for surface code cycle under depolarizing error model}
\label{table:timing_depol} 
\centering
\begin{tabular}{ | P{2.5cm} | P{3cm} |} \hline
Code distance & Avg. time / cycle \\ \hline
d=3 & 4.14msec \\ \hline
d=5 & 11.19msec \\ \hline
d=7 & 28.53msec \\ \hline
d=9 & 31.34msec \\ \hline
\end{tabular}
\end{table}

There are factors such as the number of qubits, the type of neural network being used and the number of inputs/outputs of the neural network that influence the execution time. 

Based on the analysis presented in this section, we can conclude that:
\begin{enumerate}
\item High level decoders can achieve higher decoding performance compared to low level decoders and they exhibit constant execution time with the physical error rate.
\item Recurrent neural networks reach higher decoding performance compared to Feed-forward neural networks, mainly due to the use of more weights, but that leads to a higher execution time.
\item Recurrent neural networks are much easier to train compared to Feed-forward neural networks, and have the inherent notion of time, which is beneficial for the circuit noise model. Training and evaluating a decoder based on Feed-forward neural networks becomes extremely difficult as the code distance increases.
\end{enumerate}

Based on these observations, we decided to create high level decoders based on Recurrent neural networks, in order to have a good balance between execution time and decoding performance. As we mentioned, to accomplish that balance, we search for the smallest dataset that can achieve equivalent or better decoding performance than Blossom (baseline). Increasing the number of samples in the training dataset will lead to better decoding performance at the cost of a bigger neural network requiring more training time and larger execution time. In addition, Recurrent neural networks are easier to train, allowing to decode higher code distances.

\section{Results}
\label{sec:Results}
In this section, we analyze the decoding performance of the high level decoder based on RNN for: i) different code distances, ii) the depolarizing and circuit error model, and iii) both datasets, single and multiple probabilities. 

As we previously mentioned, the way that decoding performance is tested is by running simulations that sweep a large amount of physical error rates and calculate the corresponding logical error rate for each of them. This type of simulations are frequently referred to as lifetime simulations and the logical error is calculated as the ratio of logical errors found over the error correction cycles performed to accumulate these logical errors. 
Decoding performance can also be evaluated by the pseudo-threshold for a given code distance or by the threshold for a set of several code distances for the same QEC code. The threshold represents the protection against noise for a family of error correcting codes, like the surface code. The threshold value is defined as the point of intersection of all the curves of multiple code distances (see Figures \ref{fig:depol_sing_vs_mult} and \ref{fig:circ_sing_vs_mult}).

\begin{figure}[htb]
\centering
\includegraphics[width=\columnwidth]{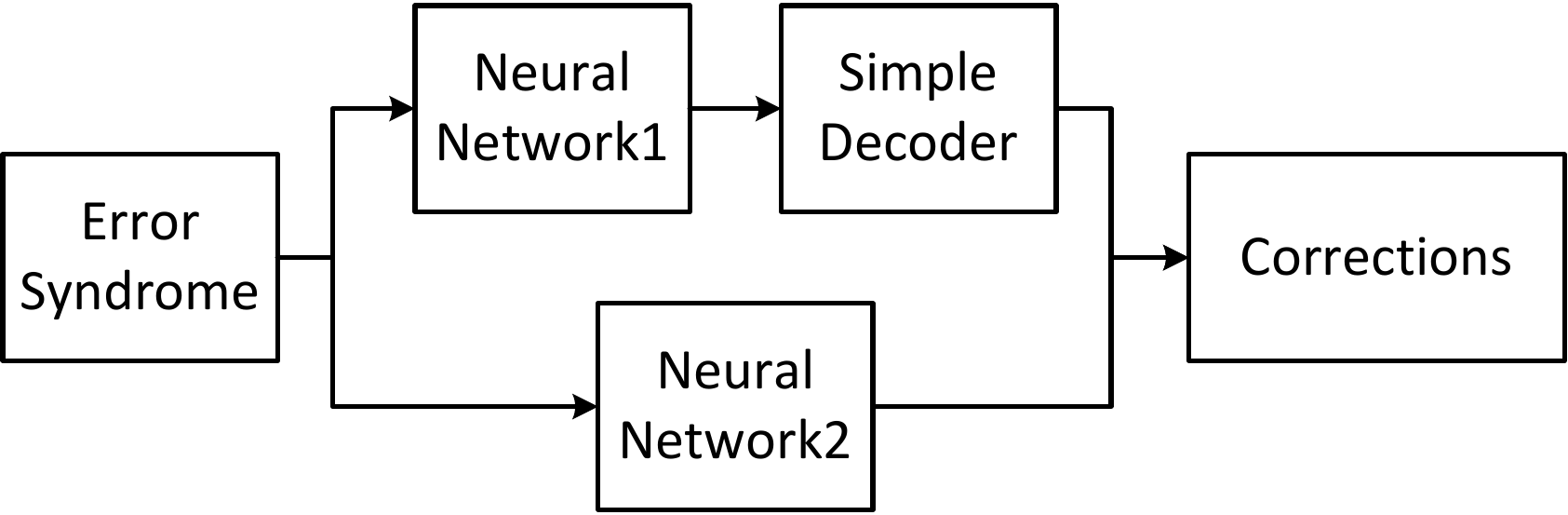}
\caption{The flow from input to output for the high level decoder that was used for the depolarizing and the circuit noise model}
\label{fig:circ_dec}
\end{figure}

The flow from input to output of the neural network based decoder that was used to obtain the results is described in Figure \ref{fig:circ_dec} for both the depolarizing and the circuit error model. For the case of the depolarizing error model, neural network 1 is not used, so the error syndrome is forwarded directly to the simple decoder since perfect syndrome measurements are assumed. The decoding process is similar to the one presented in Figure \ref{fig:pfd}. 

The decoding algorithm for the circuit noise model consists of a simple decoder and 2 neural networks. Both neural networks receive the error syndrome as input. Neural network 1 predicts which detection events at the error syndrome belong to data qubit errors and which belong to measurement errors. Then, it outputs the error syndrome relieved of the detection events that belong to measurement errors to the simple decoder. The simple decoder provides a set of corrections based on the received (updated) error syndrome. Neural network 2 receives the initial error syndrome and predicts whether the simple decoder will make a logical error and outputs a set of corrections which combine with the simple decoder corrections at the output.

\subsection{Depolarizing error model}
For the depolarizing error model, we used 5 training datasets that were sampled at these physical error rates: 0.2, 0.15, 0.1, 0.08, 0.05. Perfect error syndrome measurements are assumed, so the logical error rate can be calculated per error correction cycle.

\begin{figure}[htb]
\centering
\includegraphics[width=\columnwidth]{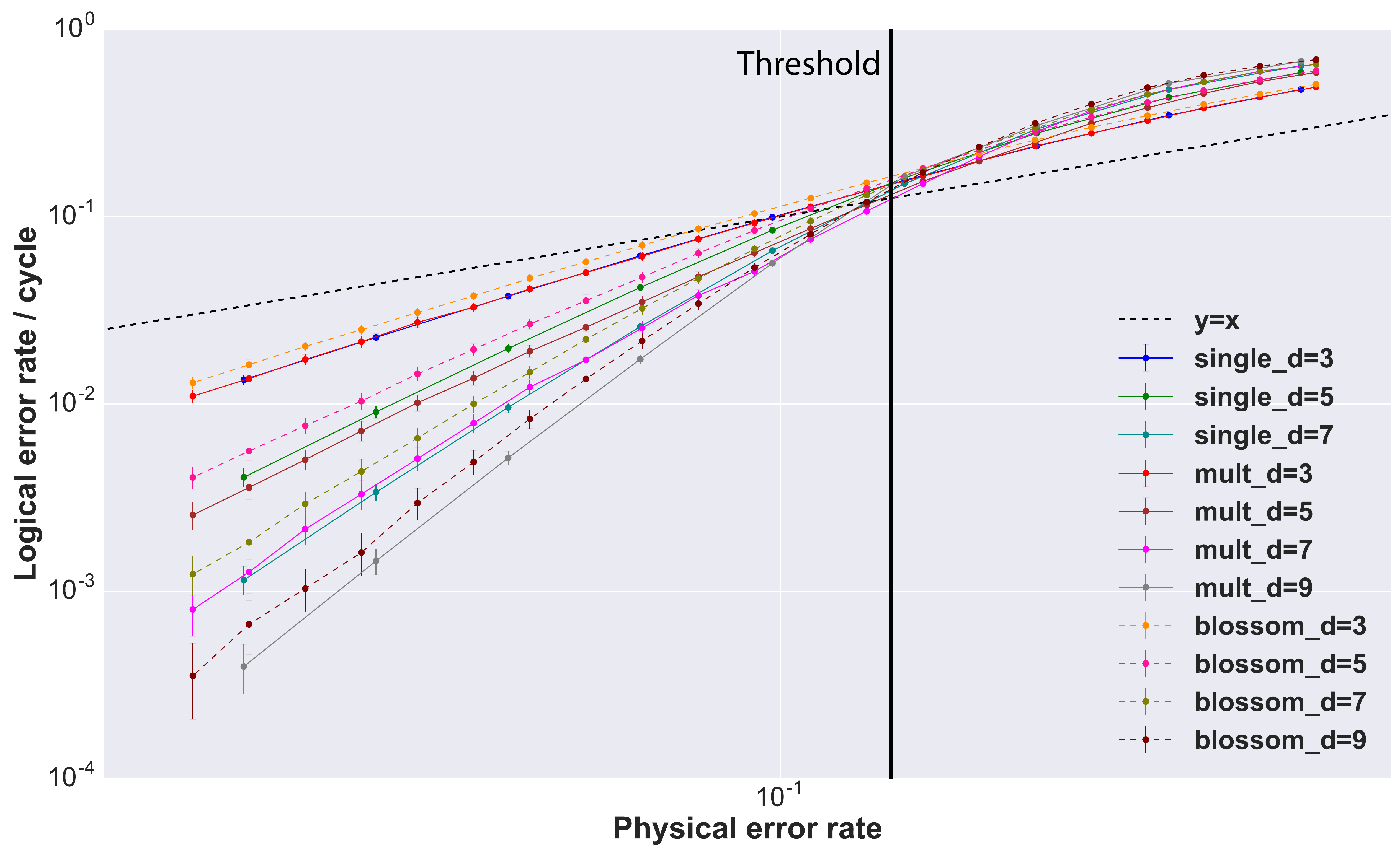}
\caption{Decoding performance comparison between the high level decoder trained on a single probability dataset, the high level decoder trained on multiple probabilities datasets and Blossom decoder for the depolarizing error model with perfect error syndrome measurements. Each point has a confidence interval of 99.9\%}\label{fig:depol_sing_vs_mult}
\end{figure}

Figure \ref{fig:depol_sing_vs_mult} shows the decoding performance achieved by the high level decoder when trained using single and multiple probabilities datsets and Blossom (baseline) under the depolarizing error model for different code distances (d= 3, 5, 7, 9). 
The neural network based decoder with the multiple probabilities datasets exhibits higher decoding performance (lower logical error rate), which is expected since it has more relevant information in its dataset. This can also be seen in Table \ref{table:pseudo_depol}, where the multiple probabilities datasets show higher pseudo-thresholds than the single probability dataset. Note that, the NN-based decoder has been trained to have similar or better decoding performance than Blossom, which is also reflected by the pseudo-threshold values.  

\begin{table}[ht]
\caption{Pseudo-threshold values for the depolarizing error model. The pseudo-threshold increases as the code distance increases}\label{table:pseudo_depol}
\centering
\begin{tabular}{ | P{2.6cm} | P{1.1cm} | P{1.1cm} | P{1.1cm} | P{1.1cm} |} \hline
Decoder & d=3 & d=5 & d=7 & d=9 \\ \hline
Blossom & 0.0823 & 0.1034 & 0.1137 & 0.1193 \\ \hline
Single prob. dataset & 0.0971 & 0.1096 & 0.1245 & N/A \\ \hline
Multiple prob. dataset & 0.0982 & 0.1219 & 0.1272 & 0.1245 \\ \hline
\end{tabular}
\end{table}

The threshold of the rotated surface code for the depolarizing model has improved from 0.140 for the single probability dataset approach to 0.146 for the multiple probabilities datasets approach, while the threshold of Blossom is calculated to be 0.142. Therefore, the goal of reaching equivalent decoding performance to Blossom was achieved. However, we should mention that the theoretical upper limit for the depolarizing error model with noiseless error syndrome measurements for the toric code is 0.189 \cite{upper}. Also, there exist various decoders that can reach close to the upper limit decoding performance for the surface code like the Maximum Likelihood Decoder (0.18) \cite{Bravyi_Suchara_Vargo}, the Markov Chain Monte Carlo (0.17) \cite{MCMC}, the neural decoder described in \cite{Krastanov} (0.165) and for the toric code the Renormalization Group decoder (0.164) \cite{RG}.

\subsection{Circuit noise model}
For the circuit noise model, we used 5 training datasets that were sampled at the following physical error rates: 4.5x$10^{-3}$, 1.5x$10^{-3}$, 8.0x$10^{-3}$, 4.5x$10^{-4}$, 2.5x$10^{-4}$. Since, imperfect error syndrome measurements are assumed the logical error rate is calculated per window of $d$ error correction cycles.

\begin{figure}[htb]
\centering
\includegraphics[width=\columnwidth]{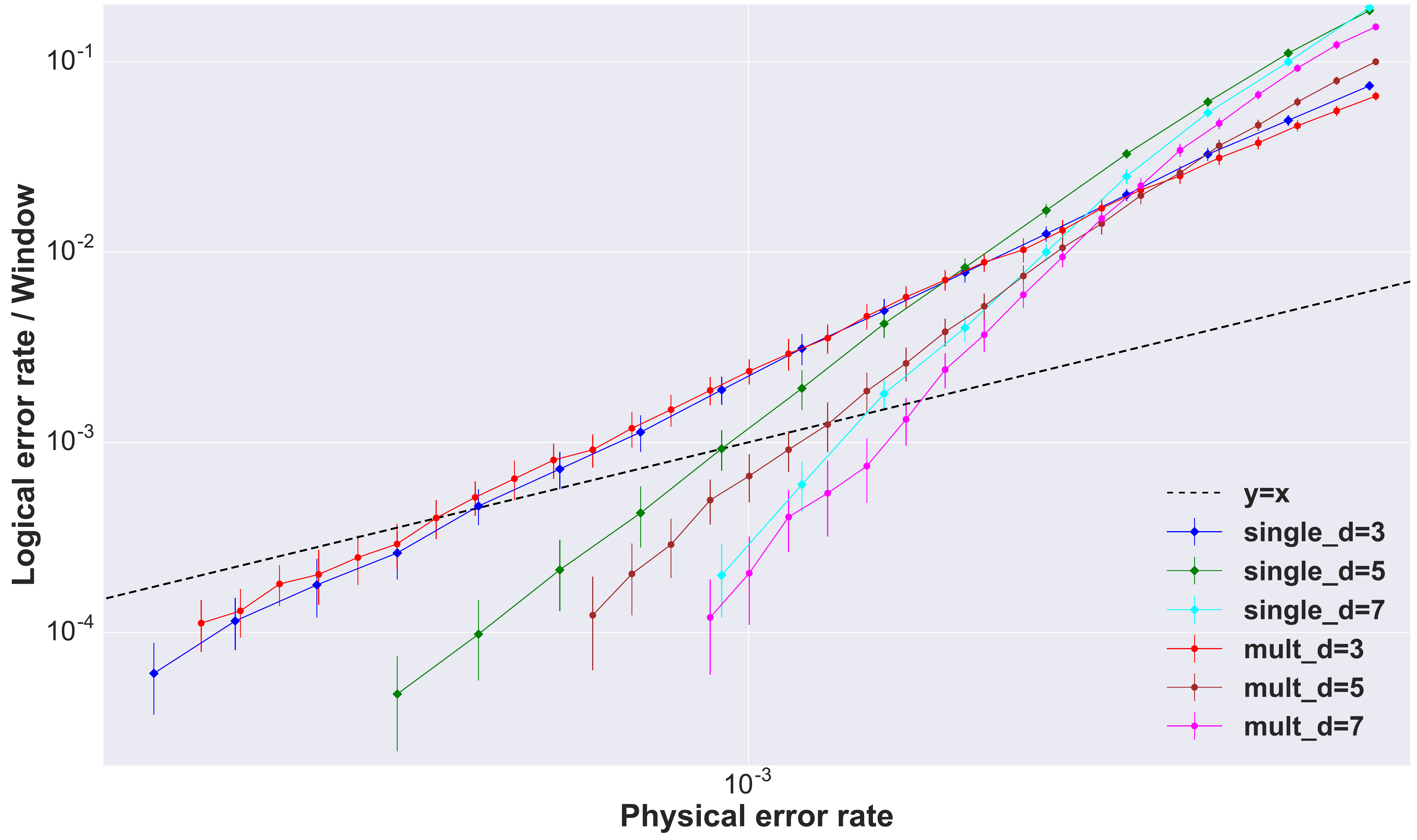}
\caption{Decoding performance comparison between the high level decoder trained on a single probability dataset and the high level decoder trained on multiple probabilities datasets for the circuit noise model with imperfect error syndrome measurements. Each point has a confidence interval of 99.9\%}\label{fig:circ_sing_vs_mult}
\end{figure}

As in the previous section, we observe from Figure \ref{fig:circ_sing_vs_mult} and Table \ref{table:pseudo_circ} that the results with the multiple probabilities datasets for the circuit noise model are significantly better, especially as the code distance is increased. The case of the d=3 is small and simple enough to be solved equally well by both approaches. In the case of d=7, we observe that for the physical error rates tested above the pseudo-threshold value, which is the sampling and training probability, the decoder does not perform efficiently, rather its performance is closer to the performance of d=5. However, below the pseudo-threshold value the samples included in the training dataset become relevant again and the performance resembles the performance of d=7.

We were not able to use the Blossom algorithm with imperfect measurements for code distances higher than 3, therefore we decided not to include it. However, we note that the results that were obtained in terms of pseudo-threshold are close to the results in the literature corresponding to the circuit noise model. The optimized version of Blossom as described in \cite{fowler_autotune, fowler_high_thres} achieves 4.20x$10^{-4}$, 2.10x$10^{-3}$ and 3.2x$10^{-3}$ for d=3, 5 and 7, respectively. Moreover, another neural network based decoder which is presented in \cite{Chamberland2018}, reports 3.18x$10^{-4}$ and 7.11x$10^{-4}$ for d=3 and d=5, respectively.

\begin{table}[ht]
\caption{Pseudo-threshold values for the circuit noise model}\label{table:pseudo_circ}
\centering
\begin{tabular}{ | P{2.4cm} | P{1.6cm} | P{1.6cm} | P{1.6cm} |} \hline
Decoder & d=3 & d=5 & d=7 \\ \hline
Single prob. dataset & 3.99x$10^{-4}$ & 9.23x$10^{-4}$ & 1.41x$10^{-3}$ \\ \hline
Multiple prob. dataset & 4.44x$10^{-4}$ & 1.12x$10^{-3}$ & 1.66x$10^{-3}$ \\ \hline
\end{tabular}
\end{table}

The threshold of the rotated surface code for the circuit noise model has improved from 2.25x$10^{-3}$ for the single probability dataset approach to 3.2x$10^{-3}$ for the multiple probabilities datasets approach, that signifies that the use of dedicated datasets when decoding a given physical error rate is highly advantageous. The surface code threshold for the Blossom algorithm is \textasciitilde 6x$10^{-3}$ \cite{fowler_autotune, fowler_high_thres}.

\section{Conclusions}
\label{sec:Conclusions}

This work focused on comparison between different decoding strategies that employ neural networks to decode the rotated surface code. Such kind of decoders are currently being investigated due to their good decoding performance and constant execution time. 

We implemented two different ways that neural networks can be used to assist (high level decoder) or perform (low level decoder) the decoding and discussed how tuning of the neural network parameters will affect their decoding performance and decoding time (execution time).

We showed that the high level decoder can achieve better decoding performance when using the same dataset size and has a constant execution time with the physical error rate for a given code distance and it increases linearly with respect to the code distance. For the FFNN and the RNN, we observed that although FFNNs can have smaller execution time, RNNs can be trained easier, therefore successfully decode higher code distances and achieve better decoding performance. Based on these observations, we further analyzed the decoding performance of a high level decoder based on a RNN.
 
We showed that sampling and training based on multiple datasets at different physical error rates, can increase the decoding performance due to the higher relevance of the samples. As expected,  equivalent decoding performance to Blossom was achieved for $d\leq 9$ for the depolarizing error model with noiseless syndrome measurements and successful decoding was achieved for $d\leq 7$ for the circuit noise model with noisy syndrome measurements.

We were not able to go to higher code distances due to the exponential increase of the dataset with the linear increase of the code distance, which poses an important challenge to such kind of decoders since much larger code distances need to be employed to have meaningful quantum computation and storage. A possible solution would be to use techniques based on distributed decoding.

\section{Acknowledgments}
\label{sec:Acknowledgments}
The authors would like to acknowledge funding from Intel Corporation.

\bibliographystyle{unsrtnat}
\bibliography{main_final}
\end{document}